\documentclass[12pt,preprint]{aastex}

\newcommand{\rosat}{{\sl ROSAT\/}}
\newcommand{\chandra}{{\sl Chandra\/}}
\newcommand{\xmm}{{\sl XMM-Newton\/}}

\newcommand{\nh}{N$_{\rm H}$}

\newcommand{\eps}{ergs\,s$^{-1}$}
\newcommand{\pscm}{cm$^{-2}$}
\newcommand{\mulx}{M101~ULX-1}

\shorttitle{X-ray Properties of \mulx}
\shortauthors{Mukai et al.}

\begin{document}


\title{The X-ray Properties of \mulx = CXOKM101~J140332.74+542102}


\author{K. Mukai\altaffilmark{1}, M. Still\altaffilmark{1}, R.H.D. Corbet\altaffilmark{1}}
\affil{Exploration of the Universe Division,
       Code 662, NASA/Goddard Space Flight Center, Greenbelt, MD 20771}
\email{mukai@milkyway.gsfc.nasa.gov}
\author{K.D. Kuntz}
\affil{Henry A. Rowland Department of Physics and Astronomy,
       Johns Hopkins University, Homewood Campus, Baltimore, MD 21218,
       and Exploration of the Universe Division,
       Code 662, NASA/Goddard Space Flight Center, Greenbelt, MD 20771}

\and
\author{R. Barnard}
\affil{Department of Physics and Astronomy, The Open University
       Walton Hall, Milton Keynes MK7 6BT, UK}


\altaffiltext{1}{Also Universities Space Research Association}


\begin{abstract}

We report our analysis of X-ray data on \mulx, concentrating on
high state \chandra\ and \xmm\ observations.  We find that the
high state of \mulx\ may have a preferred recurrence timescale.
If so, the underlying clock may have periods around 160 or 190 days,
or possibly around 45 days.  Its short-term variations resemble
those of X-ray binaries at high accretion rate.  If this analogy
is correct, we infer that the accretor is a 20--40 M$_\odot$ object.
This is consistent with our spectral analysis of the high state
spectra of \mulx, from which we find no evidence for an extreme
($> 10^{40}$ \eps) luminosity.  We present our interpretation in
the framework of a high mass X-ray binary system consisting of a
B supergiant mass donor and a large stellar-mass black hole.

\end{abstract}




\section{Introduction}

It is now clear that some X-ray sources in external galaxies are
brighter than those seen in our Galaxy, thus providing a strong
scientific motivation for high spatial resolution X-ray studies.
Off-nuclear sources in nearby galaxies with luminosities
exceeding 10$^{39}$ \eps\ have been designated ultraluminous X-ray
sources (ULXs).  The nature of ULXs, in particular whether they are
powered by intermediate-mass black holes (IMBH, M$>$100 M$_\odot$)
that are distinct from stellar-mass and supermassive black holes, has
become a hotly debated issue \citep{F2005}.  If their X-ray emissions
are isotropic and sub-Eddington, the compact objects in ULXs must be
at least 7 solar masses (7M$_\odot$).

Our object of study here is CXOKM101~J140332.74+542102 (hereafter \mulx),
a highly variable source in the nearby (7.2 Mpc; \citealt{Sea1998})
face-on spiral galaxy, M101, whose ULX phase was discovered in the
2000 March \chandra\ observations \citep{Pea2001,Mea2003}.
\citet{Kea2005} have discovered an optical counterpart of \mulx,
with brightness and colors consistent with those of a B supergiant
in M101, hence it is almost certainly a high-mass X-ray binary (HMXB).
Also throughout 2004, we carried out a series of \chandra\ observations
totaling a million seconds of exposure time (hereafter ``Msec'') for a
comprehensive X-ray study of M101, including monitoring of bright individual
sources.  \citet{KDY2004} have analyzed a subset of these data and claim
that \mulx\ reached a peak 0.3--7 keV luminosity of 3$\times 10^{40}$
\eps\ in July 2004, a claim which we reexamine in \S 3.3, after considering
the long-term and short-term variability in \S 3.1 and \S 3.2, respectively.
We then reassess the nature of \mulx\ in \S 4.

\section{Observations}

The Msec data set, as well as our reduction method, will be
described fully in \citet{K2005}.  In addition, we have
analyzed all archival \rosat, \chandra, and \xmm\ observations
of M101 to establish the long-term history of \mulx.  Here
we present our detailed analysis of the high state \chandra\ and
\xmm\ data.  The relevant datasets are summarized in Table\,\ref{hsdata}.
We do not analyze the low state data in detail; any effort to
analyze them (e.g., \citealt{KDY2004}) is necessarily limited
by the poor statistics.  In addition, there is potential for source
confusion in a low state, since we cannot rule out a faint contaminating
source at luminosities up to 10$^{36}$ \eps.

\section{Analysis}

\subsection{Long-term Variability}

\begin{figure}
\plotone{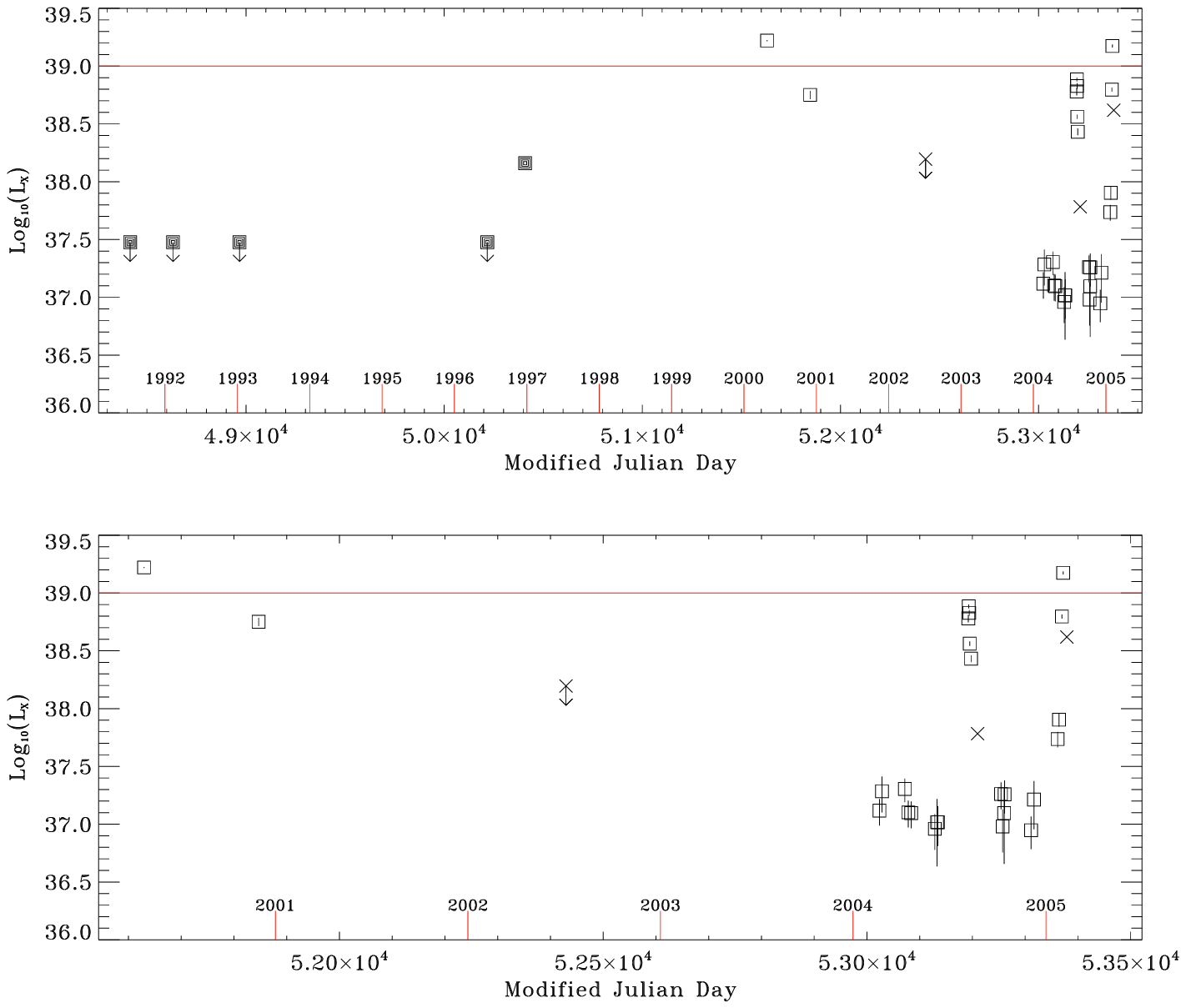}
\caption{The long-term light curve (0.3--2.5 keV luminosity)
of \mulx\ from \rosat\ (filled square), \chandra\ (open square),
and \xmm\ (cross) data, in two panels.  The upper panel shows
the light curve over the period 1991--2005, while the lower panel
shows a blow up for the period 2001--2005.}
\label{ltlc}
\end{figure}

We have reconstructed the long-term light curve of \mulx.  We compute
the 0.3--2.5 keV luminosity using an absorbed blackbody plus power law
model for the high signal-to-noise data (all data marked as ``high state''
in Table 1 as well as the 2000 October \chandra\ and 2004 July \xmm\ 
observations).  For fainter data, we convert from count rates to luminosity
using a conversion factor derived from 2004 July \xmm\ data.  We similarly
estimate the luminosity during earlier \rosat\ observations.  We
present the resulting light curve in Figure\,\ref{ltlc}, once for the
entire period (1991 June through 2005 January) and once for the
\chandra-\xmm\ era (2000 March through 2005 January).  The most obvious feature
of this is the recurrent high states. We investigate if there is an
underlying periodicity in these high states.

High states of \mulx\ appear to have durations of over eighteen days,
though the amount of time spent at the peak luminosity is significantly
shorter, judging from the lightcurves of the 2004 July and 2004
December/2005 January high states.  The intervals between the 2000 March
high state and the 2000 October high state (216 days) and that between the
2004 July high state and the 2004 December/2005 January high state (151
days) are incommensurate, implying that the high states of \mulx\ does
not repeat on a strict period.

In order to explore the characteristic time scales of the system,
we have classified each observation as either ``high'' (L$_X>10^{37.5}$ \eps)
or ``low.''  We attempted to match this bimodal light-curve
with a square wave with period $P$, phase, and an activity fraction
as free parameters.
If the source were strictly periodic,  then the activity fraction would
correspond to the duty fraction. For a source that is not strictly periodic,
the activity fraction corresponds to the width
of a ``window'' in which the source is likely to be active,
convolved with the duty fraction.

We have explored three different data selections: The well sampled
interval from 2004 January to 2005 January (the Msec interval),
the less well sampled interval from 2000 March to 2005 January
(the \chandra\-\xmm\ era), and the entire interval, which includes
the \rosat\ data as well.  These selections allow us to investigate
a range of periods, allowing for departures from a strict periodicity.

We have tested periods from 20 to 250 days, and activity fractions
from 1\% to 60\%.  Figure~\ref{periods} (left) shows the regions
with the best matches to the data for the three different data selections.
Note that, given the high/low simplification of the light curve used
in our analysis, the regions in this plot are either allowed or not
allowed, with no quantitative goodness of fit in the usual sense.  The
Msec epoch data allow ``periods'' of 154--163 days and 187--199 days, with
no data-model mismatches.  The more extended interval does not allow
a solution if we require that all data points match the model.  However,
if we allow one mismatch, periods of 159 days and 189--199 days are
allowed. The entire dataset has a periodic solution only
if we allow two observations to be mismatched. ``Periods'' with only two
mismatches are 156 days, 159 days, 172--174 days, and 193--197 days.
Thus, no matter which data subset was chosen, ``periods'' around 159
and 195 days best matched the data, but the longer the interval,
the more mismatches are required.

We also investigated if a shorter period ($<$60 days) might
match the data (Figure~\ref{periods} right).  Using the Msec
interval, the best solutions (45 days)
would produce two mismatches, usually the 2004 July \xmm\ observation,
and either the first or last observations of an extended set of
\chandra\ observations.  Longer fit intervals produced greater
numbers of mismatches for their best fits.

This analysis suggests that \mulx\ has a characteristic time-scale.
It is not a strict periodicity, although it is still possible that
it is a superposition of periodic and random variations.

\subsection{Short-term Variability}

We present the light curves of 2004 July and 2004 December/2005
January high states in Figure\,\ref{lcjuly} and Figure\,\ref{lcjan},
respectively.  The high degree of short-timescale variability is
obvious.  Moreover, we confirm our earlier results \citep{Mea2003}
of the energy dependence of the variability: \mulx\ is more variable
at higher energies.

The power density spectra (PDS) of disk-accreting X-ray binaries
can be classified into two broad categories based on whether
the Eddington accretion rate ($l$=$L$/$L_{\rm Edd}$) is above
or below the critical value, $l_c$, regardless of the nature
of the primary \citep{vdK1994}.  At low accretion rate ($l < l_c$),
X-ray binaries have a fractional rms amplitude of a few times 10\%
and a broken power law PDS (hereafter Type A); at high accretion
rate ($l > l_c$), they have a fractional rms of a few percent, with
power law (index 1--1.5) PDS (Type B).  Although there is a well-known
hysteresis effect \citep{Mea1995} that complicates the situation,
\citet{Bea2004} have nevertheless obtained an empirical calibration
of $l_c$ of 0.1, in terms of observed 0.3--10 keV luminosity as a
fraction of the Eddington limit.  They (see also \citealt{Bea2003})
have applied this to identify black hole candidates in M31.

We have performed a similar PDS analysis of \mulx, both
from the 2000 March observation  and from the 2004 July
observations (Figure\,\ref{pdslong}).  The PDSs can
be described by a power law with a slope of about $-$1, merging into
the photon counting noise at 10$^{-3}$ Hz.  This is consistent
with a Type B PDS for an extragalactic X-ray binary, implying that
the Eddington accretion rate of the source was in the range
0.1--1.0 both in 2000 March and in 2004 July.  Therefore, equating
our estimate of the 0.3--10 keV luminosity in 2004 July of
$\sim 6 \times 10^{38}$ \eps\ (\S 3.3) with 0.1 L$_{\rm Edd}$,
we infer the upper limit of compact object mass of $\sim$ 40 M$_\sun$.
If the X-rays we observe from \mulx\ are beamed, then the true luminosity
is lower, and the compact object mass is also lower.  On the other hand,
equating the highest luminosity we measure (3$\times 10^{39}$ \eps) with
L$_{\rm Edd}$, we infer a lower limit of 20 M$_\sun$, again assuming that
there is no beaming and \mulx\ stays sub-Eddington at all times.  However,
the latter assumption need not apply strictly, since it is quite possible
for an accreting object to exceed its Eddington limit for a short time.

\subsection{High State Spectra}

We have investigated the high state spectra of \mulx, first using
an absorbed blackbody model \citep{Mea2003,KDY2004}.  When applied
to the 2004 July data, we obtain an high absorbing column
(\nh $\sim 2 \times 10^{21}$ \pscm) and an absorbed 0.3--7 keV
luminosity of $\sim 4 \times 10^{38}$ \eps.  (Note that the
3$\times 10^{40}$ \eps\ value claimed by \citet{KDY2004} is the unabsorbed
luminosity in the 0.3--7 keV band).

The X-ray absorbing column inferred by this fit is well above the
Galactic column of $\sim 10^{20}$ \pscm\ and the total absorption
through the disk of M101 ($\sim 6 \times 10^{20}$ \pscm).  If this
X-ray column was due to normal interstellar matter, this translates
to an expected color excess E$_{B-V}$ of $\sim$0.33 \citep{Bea1978},
or a visual extinction A$_V$ of $\sim$1 magnitude.  In contrast, the
optical counterpart has a B$-$V color of $-$0.15$\pm$0.12,
which is consistent with that of a B supergiant with little reddening
(see Figure 4 of \citealt{Kea2005}).  In fact, an assumed value of
A$_V$=0.4 will make the optical counterpart intrinsically bluer than
any early type stars.  Moreover, there is a suggestion of day-to-day
variability of \nh\ during the interval July 5--12, 2004 in the spectral
fits by \citet{KDY2004}.  Therefore, if this model is correct, a significant
fraction of the X-ray absorber is likely to be located within the binary
or in its immediate surroundings and is thus subject to an intense
ionizing radiation.

Given this, we have also fitted the average (July 5--11) spectrum
using a blackbody plus ionized absorber model (the latter implemented
in XSPEC as ``absori'').  Highly ionized matter is essentially transparent
at soft X-ray energies, and even a small ($\xi = 1$) ionization makes
the fit visibly worse (Figure\,\ref{absori}; Table\,\ref{absfit}). 
We therefore conclude that the observed shape of the soft X-ray spectrum
demands a near neutral absorber, if a blackbody (or a disk blackbody) is the
intrinsic spectrum shape.  What if the intrinsic spectral shape is a
blackbody with absorption edges in the \chandra\ band, as \citet{KDY2004}
claimed?  We can indeed fit the 2004 July spectrum with modest interstellar
column (4$\times 10^{20}$ cm$^{-2}$), but only if we include a strong edge
below the ACIS band, e.g., at 0.28 keV.  We show one such example in
Figure\,\ref{absori}, resulting in a 63 eV blackbody with edges at
0.28 and 0.56 keV.  However, a strong edge below 0.3 keV requires
low Z elements (presumably C) with K-shell electrons.  That is, such
an edge is in fact the sign of a near neutral absorber.  We therefore
conclude that the presence of a neutral absorber is required to explain
the observed curvature of the spectrum below $\sim$0.5 keV.

We have then analyzed all high state spectra obtained with
\chandra\ together.  These are the 2000 March
observation analyzed by \citet{Mea2003}; the 2004 July observations;
and the 2004 December/2005 January observations.  We omit the 2000 October
observation, since \mulx\ was relatively faint, although still
well within the high state, and the exposure time was short ($<$10 ksec).
We re-use the subdivision of 2000 March data into high, medium, and low
intensity states as defined by \citet{Mea2003}.  When \mulx\ is at
its brightest, we observe an excess of counts above 2 keV,
which can be fitted with an additional power law.  However, the parameters
of this component is poorly constrained, so we concentrate on fitting
below 2 keV.

Using the blackbody model absorbed only with a neutral absorber, we plot
the blackbody temperature, the absorbing column, and the inferred bolometric
luminosity against the count rates in the left column of Figure\,\ref{trend}.
We use the observed count rates for the Msec observations, but scale
the 2000 March count rates by the ratio of 0.3--1 keV effective areas to
approximately account for the loss of low energy response in ACIS-S.

We immediately see that both N$_H$ and the inferred bolometric
luminosity (L$_{\rm bol}$) have a negative correlation with
the observed count rate.  The similarity of behavior of the
two suggests there is a common origin.  We propose that this
is an artifact of the absorbed blackbody fit: if the true
spectrum of \mulx\ is more strongly peaked than a blackbody, blackbody fits
will forced to a high N$_H$ value and inflate the inferred L$_{\rm bol}$.

To test this hypothesis, we have fitted these 6 spectra with
a second model.  After some experimentation, we have fixed N$_H$ to
4$\times 10^{20}$ cm$^{-2}$, a reasonable value for an object in
the disk of M101.  The underlying model is a variable blackbody
plus a relativistic line from an accretion disk (``diskline'' as implemented
in {\tt xspec})\footnote{A simple Gaussian line is a valid alternative.
However, the asymmetric shape of the ``diskline'' profile with sharp
drop of the blue wing and a more gradual drop of the red wing gives
a somewhat better fit for some of the spectra.}.  This model is
often invoked for the Fe K lines in active galactic nuclei (AGN).  More
recently, \cite{Bea2001} have fitted the \xmm\ RGS spectra of two AGN
using relativistic lines from OVIII, NVI, and CV, and the same physical
interpretation can in principle apply to \mulx.

The parameters for the diskline were fixed at the following values,
after experimentation: the central energy, 0.5 keV; $\beta$ (power
law index for the radial dependence of emissivity), $-$2; range of
accretion disk radius, 6--1000 M$_g$; and disk inclination, 75$^\circ$.
We show the July 2004 data as fitted with this model in Figure\,\ref{absori}.
The line energy (0.5 keV) suggests NVI, while for this spectrum alone
(in fact, only the July 6 spectrum), an additional line at 0.8 keV would
further improve the fit.  Alternatively, this feature could be due to a
transient edge at 0.9 keV \citep{KDY2004}.  As for the 0.6 keV edge also
noted by \cite{KDY2004}, this is not required for a good fit once the
0.5 keV emission line is included in the model.

We summarize the results for all
high state spectra in Table\,\ref{trendtab} and in the right
column of Figure\,\ref{trend}.  In contrast to the absorbed blackbody
fits, the inferred L$_{\rm bol}$ values are positively correlated with
the count rates  using this model.    The inferred bolometric
luminosity and the unabsorbed 0.3--10 keV luminosity for the 2004 July
data are $\sim 1 \times 10^{39}$ \eps\ and $\sim 6 \times 10^{38}$ \eps,
in contrast to the much higher value obtained with the absorbed
blackbody fit.  The 2005 January \xmm\ data can
also be fitted with either model (Table\,\ref{trendtab}).  In these
fits, the model parameters are statistically well-constrained  (e.g.,
the blackbody temperature has typical errors of 5 to 20 eV), but it is
clear that the true uncertainties are dominated by the systematics
due to our choice of models.

\section{Discussion}

\subsection{Nomenclature}

Various authors have borrowed existing terminology to describe
\mulx\ and other bright X-ray sources in nearby galaxies.
On closer inspection, however, the applied terminology often
fails to meet the traditional definition used in the Galactic
X-ray Binary community.  We have therefore considered, and ultimately
declined to use, two nomenclatures that have been applied to \mulx.

Super-soft sources (SSSs) are luminous soft X-ray sources with
equivalent blackbody temperatures of $\sim$15--80 eV and estimated
bolometric luminosities of 10$^{36}$--10$^{38}$ \eps\ \citep{KvdH1997}.
It is generally accepted that, with few exceptions, SSSs are nuclear
burning white dwarfs.  However, some authors (see, e.g., \citealt{dSK2003})
have sought to expand the definition of the class in recent years.
While the expanded definition allows these authors to draw attention
to a collection of interesting objects, they are likely to be
a heterogeneous mixture.  We prefer to preserve the well-established
connotation of SSS with nuclear burning white dwarfs.  \citet{Pea2001}
identified 10 \chandra\ sources in M101 as supersoft, using a hardness
ratio criterion that approximates the traditional definition: these
sources have no counts above 0.8 keV.  On the other hand, \mulx\ was
in a high state at the time and had significant counts even above 1.3 keV.
We therefore do not consider \mulx\ to be an SSS.

The other terminology we discuss is ``X-ray transient,'' for which
it is difficult to find a formal definition.  Historically,
this term is applied to Galactic X-ray sources that are detectable
by all-sky monitors at one epoch and undetectable in pointed observations
using collimated (non-imaging) instruments.  In particular, soft X-ray
transients, which are believed to be due to disk instability \citep{L2001},
usually have quiescent luminosity less than 10$^{34}$ \eps.  With the
current generation of instruments, sources in M101 that are fainter than
10$^{36}$ \eps\ are undetectable in a single observation.  Such an upper
limit is far too high to establish the presence of a true transient,
although it is certainly possible to find candidate transients \citep{Jea2005}.
In the specific case of \mulx, it is unknown if it ever becomes
fainter than 10$^{36}$ \eps.  For this reason, we have not used the
term ``transient'' to describe \mulx, nor have we used the related terms
``quiescent'' and  ``outburst,''  instead preferring the more neutral
terms, ``high state'' and ``low state.''

\subsection{The Nature of the Compact Object}

\citet{KDY2004} have analyzed the 2004 July spectra of \mulx\ using
the absorbed blackbody model, inferred a bolometric luminosity of
$\sim 10^{41}$ \eps, and concluded that \mulx\ must contain an IMBH.
Using the Eddington luminosity argument, $\sim 10^{41}$ \eps\ corresponds
to a minimum mass of 700 M$_\odot$.  However, our spectral analysis
above makes it clear that this conclusion is highly model-dependent.
We argue, moreover, that the absorbed blackbody fits have two
shortcomings: the requirement for a neutral absorber next to a
$>$10$^{40}$ \eps\ X-ray source, and the counterintuitive trend
of fitted parameters against \chandra\ count rate.

As noted earlier, the high and variable values of the X-ray absorber
in the absorbed blackbody fits (Figure\,\ref{trend}) require the
absorber to be in the vicinity of \mulx\ itself.  Whereas the presence
of such intra- or circum-binary
material is expected, due to the copious mass loss from the supergiant
mass donor, it is likely to be highly ionized on average (see, e.g.,
\citealt{Wea2000}), even when the luminosity is a few times 10$^{38}$ \eps;
much more so if the bolometric luminosity is near 10$^{41}$ \eps.  While
high density clumps with lower than average ionization are expected, we
need extreme clumping to explain the observed absorber.  For example,
for a clump 10$^{13}$ cm from a 10$^{41}$ \eps\ source to have $\xi < 1$,
a density greater than 10$^{15}$ cm$^{-3}$ is required; for such a clump
to be responsible for the observed X-ray absorption of
\nh\ $\sim 2 \times 10^{21}$ \pscm,
it must have a length of only $\sim 2 \times 10^6$ cm.  A clump at
a distance of 10$^{15}$ cm needs a density $> 10^{11}$ cm$^{-3}$
and a length of $\sim 2 \times 10^{10}$ cm.  That is, for such a clump
to remain neutral, it must have a high overdensity factor, and hence
a small filling factor.  We therefore consider the inferred neutrality
of the X-ray absorber to be a severe problem for the absorbed blackbody
interpretation.

This model also leads to a counterintuitive trend with observed
count rate (Figure\,\ref{trend}; Table\,\ref{trendtab}), whereas
our alternative spectral model does not suffer from this problem.
We do not necessarily claim that our model is the correct physical
description of \mulx.  Nevertheless, it is comforting to note that
a simple analytical model, with a physical interpretation (relativistic
emission line from the accretion disk) can fit the collection of high state
spectra without requiring an anti-correlation between observed
count rate and bolometric luminosity, or the presence of a neutral
absorber within an ULX system.  It is quite possible that other models
can be found that can fit the available data equally well.  Any model
that fits the data without requiring a high \nh\ will likely result
in a moderate bolometric luminosity in 2004 July of
$\sim 1 \times 10^{39}$ \eps.

We conclude that there is no compelling evidence for an IMBH
of the sort claimed by \citet{KDY2004} ($>>$100 M$_\odot$) in \mulx.
We furthermore consider that the accreting object is likely
a 20--40 M$_\sun$ black hole from our PDS analysis (\S 3.2).
The X-ray spectrum of \mulx\ is unlike those of well-established
stellar-mass black hole candidates, but this is common problem for
all models of \mulx.  Before proceeding further with this interpretation,
we briefly consider possible alternatives.

Can \mulx\ contain a nuclear burning white dwarf, given that the high state
spectrum of \mulx\ is dominated by a soft, blackbody-like component? 
There is indeed a superficial similarity between \mulx\ and SSS.  However,
\mulx\ is too hot at times (kT as high as 0.2 keV), and the observed
luminosity in 2000 March (before absorption and bolometric corrections)
is well above the Eddington limit for a white dwarf.  High luminosity is
thought to drive a strong (but non-relativistic) outflow, expanding
the white dwarf photosphere and driving the emission to longer wavelength
(see, e.g., \citealt{K1997}).  Beaming is unlikely, since the supersoft
emission in a nuclear burning white dwarf is photospheric, and no
relativistic jet is expected.  Thus, a white dwarf accretor appears
unlikely in this case.

Can it be a neutron star?  Many X-ray pulsars have a soft component often
modeled as a blackbody with kT$\sim$0.1 keV \citep{Hea2004},
which is also pulsed.  If it is a pulsar, its emission is by definition
beamed, so the Eddington luminosity argument loses its punch.  Although
accreting pulsars are usually dominated by a hard power-law component
in the \chandra\ band, different beaming of the hard and soft component
might make \mulx\ appear dominated by the soft component.  Our timing
analysis is limited by the time resolution of the ACIS data in the high
state and the lack of photons in the low state, so a P$_{spin} \sim 1$s
pulsar would have not been detected.  This possibility should be kept
in mind for \mulx, but does require a factor of 10 beaming of the
soft component to explain the highest luminosity data.

\subsection{The Nature of the Binary System}

We therefore return to the interpretation we consider most likely,
that \mulx\ is a high-mass X-ray binary (HMXB) with a B supergiant
mass donor \citep{Kea2005} and a 20--40 M$_\sun$ black hole.  At the
low end of our preferred range, it can be considered a normal stellar-mass
black hole, while at the high end, it is more massive than those
predicted by the standard theory of stellar evolution.  It is unclear,
however, if a radically different formation mechanism is required
for such a black hole.  In any case, the binary as a whole shows
a family resemblance to the more familiar HMXBs.

One well-known system to which \mulx\ might be compared is Cyg X-1,
for which the transitions to and from the low/hard state and the
high/soft state enhance the apparent variability, depending on
the bandpass of the X-ray instrument.  Its count rates varies by
a factor of $\sim$10 in the {\sl RXTE\/} ASM, which is sensitive to 2--10 keV
X-rays.  With the exception of the orbital period, there is no known
periodic component in its long-term light curves, even though
there are well-studied correlations between average brightness,
X-ray spectral shape, X-ray PDS, and the radio emission \citep{Gea2004a,
Gea2004b}.  We might simply consider \mulx\ to be an extreme
cousin of Cyg X-1, perhaps related to the higher black hole mass.

Alternatively, the long-term modulation of \mulx\  may be a ``super-orbital''
variability seen in several Galactic and Magellanic X-ray binaries
(see, e.g., \citealt{Cea2004}).  Radiation-induced warping of
the accretion disk is the leading candidate mechanism for such
modulations; it typically operates in high-luminosity, Roche-lobe
overflow systems and the resultant modulations are not strictly
periodic.

\begin{figure}
\plottwo{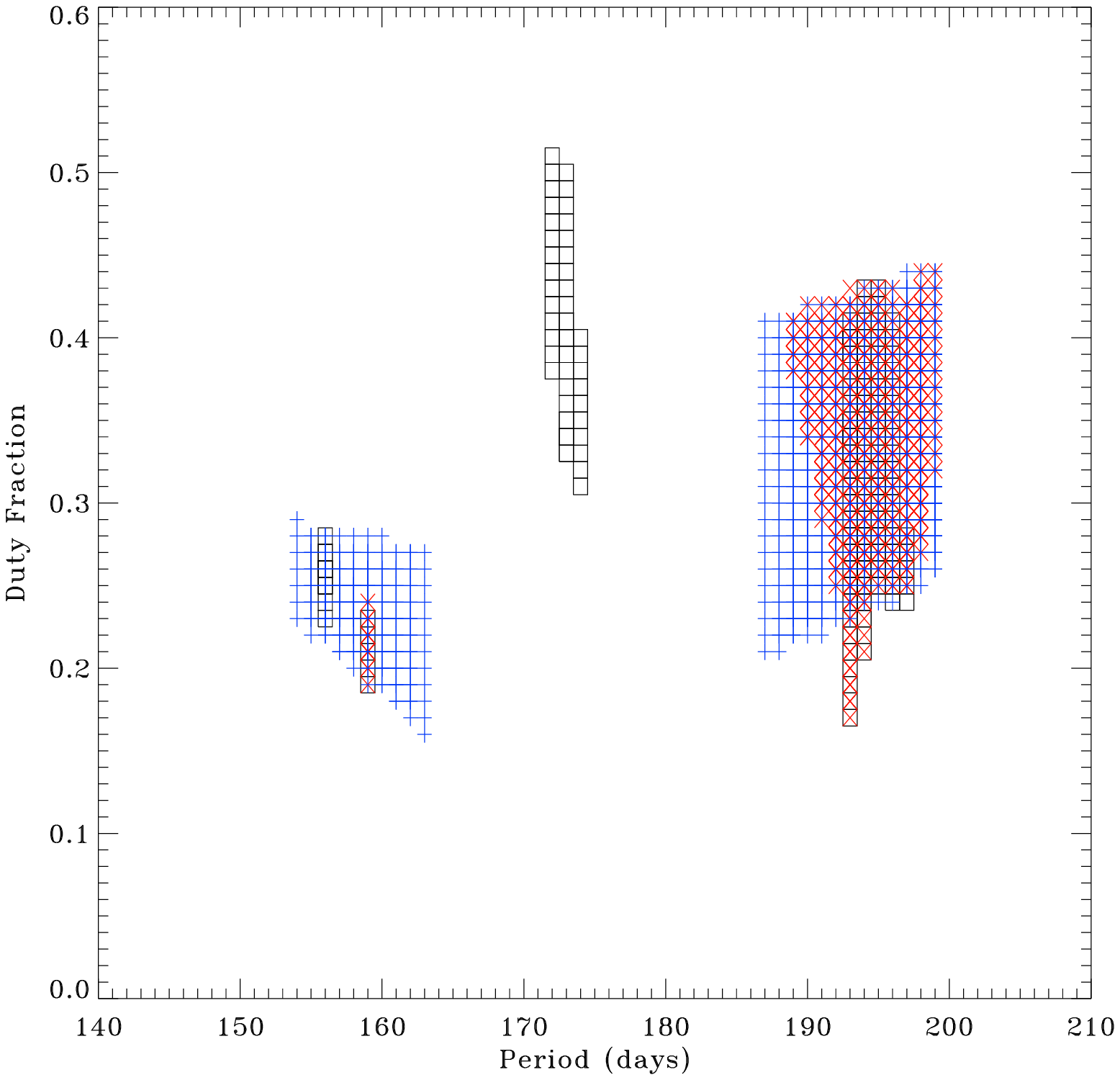}{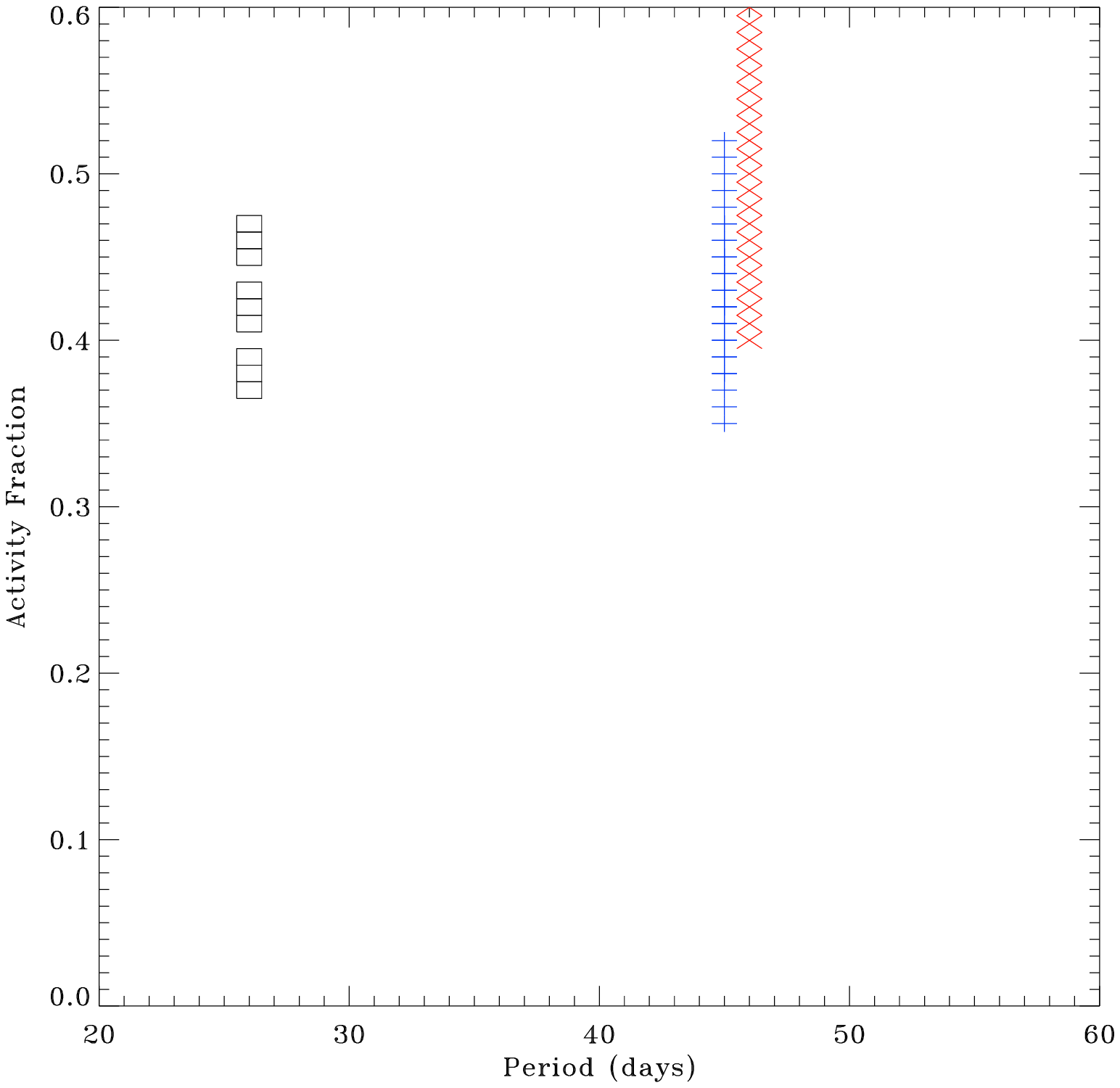}
\caption{{\bf Left:} The region of the period-activity space
with the best match to the data using the Msec interval 
({\it Crosses} no mismatches), 
the Chandra-XMM era interval ({\it Xs} one mismatch), 
and the entire interval ({\it Squares} two mismatches).
{\bf Right:} 
The short-period region of the period-activity space
with the best match to the data using the Msec interval 
({\it Crosses} two mismatches),       
the Chandra-XMM era interval ({\it Xs} three mismatch),                 
and the entire interval ({\it Squares} four mismatches).}
\label{periods}
\end{figure}

\begin{figure}
\plotone{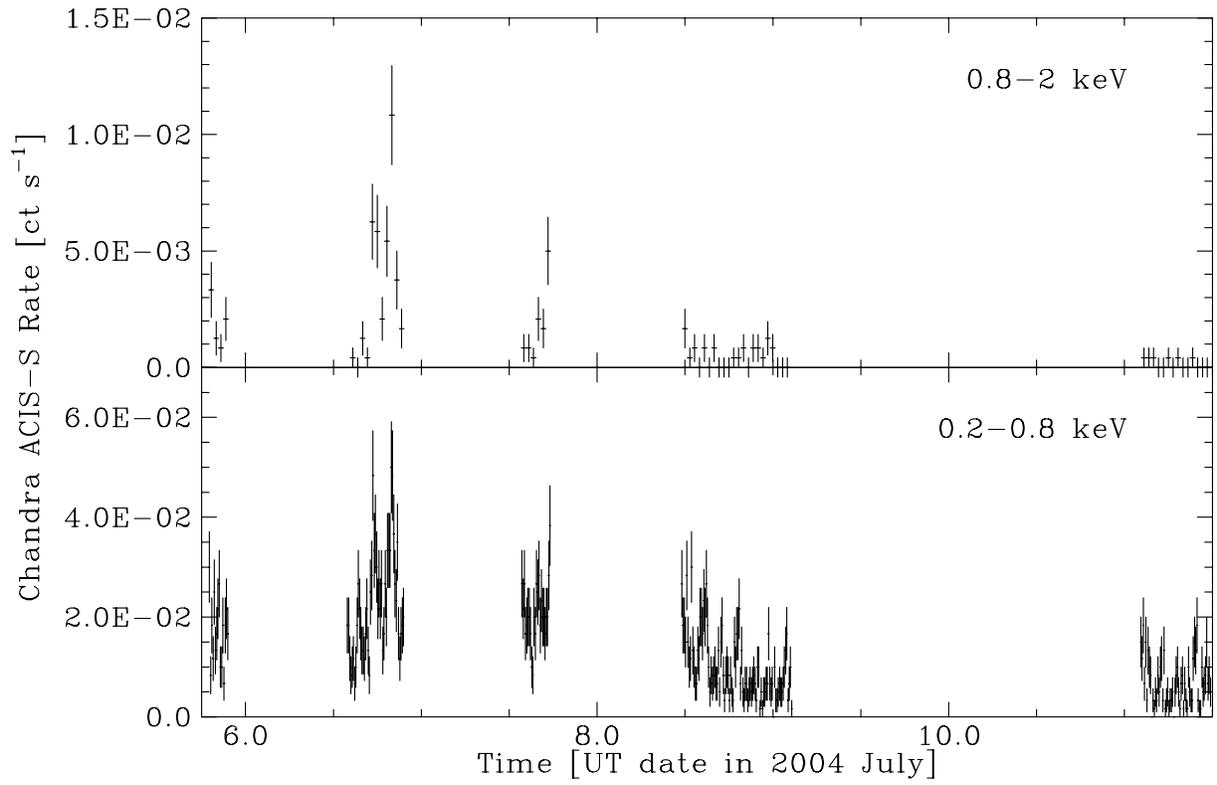}
\caption{\chandra\ ACIS-S light curve of \mulx\ in 2004 July
in two bands; one bin is 600s for the 0.2--0.8 keV band and
2400s for the 0.8--2 keV band.}
\label{lcjuly}
\end{figure}

\begin{figure}
\plotone{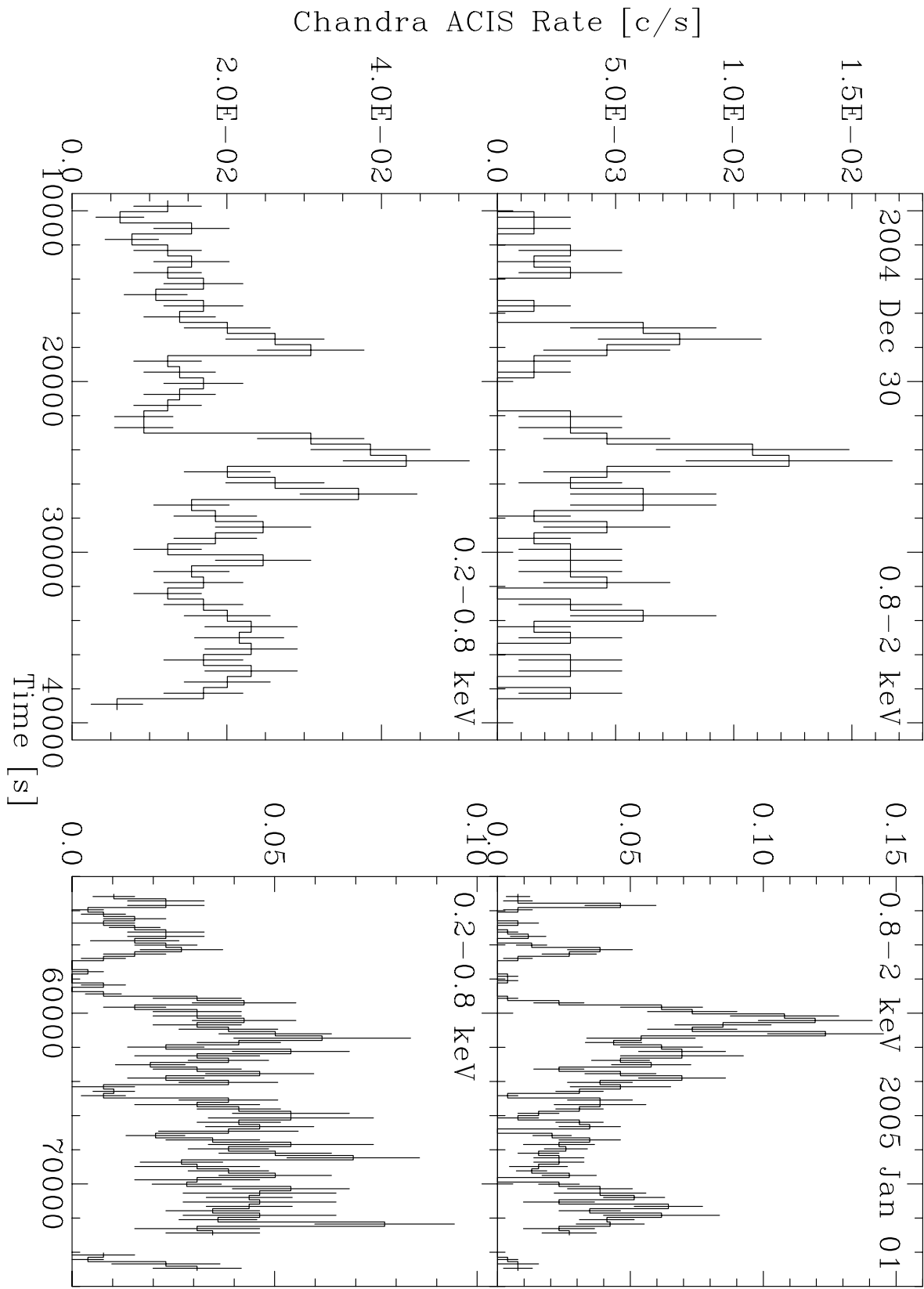}
\caption{\chandra\ ACIS-S light curve of \mulx\ on 2004 December 30
(in 648 s bins) and on 2005 January 1 (256 s bins), in two energy
bins.}
\label{lcjan}
\end{figure}

\begin{figure}
\plotone{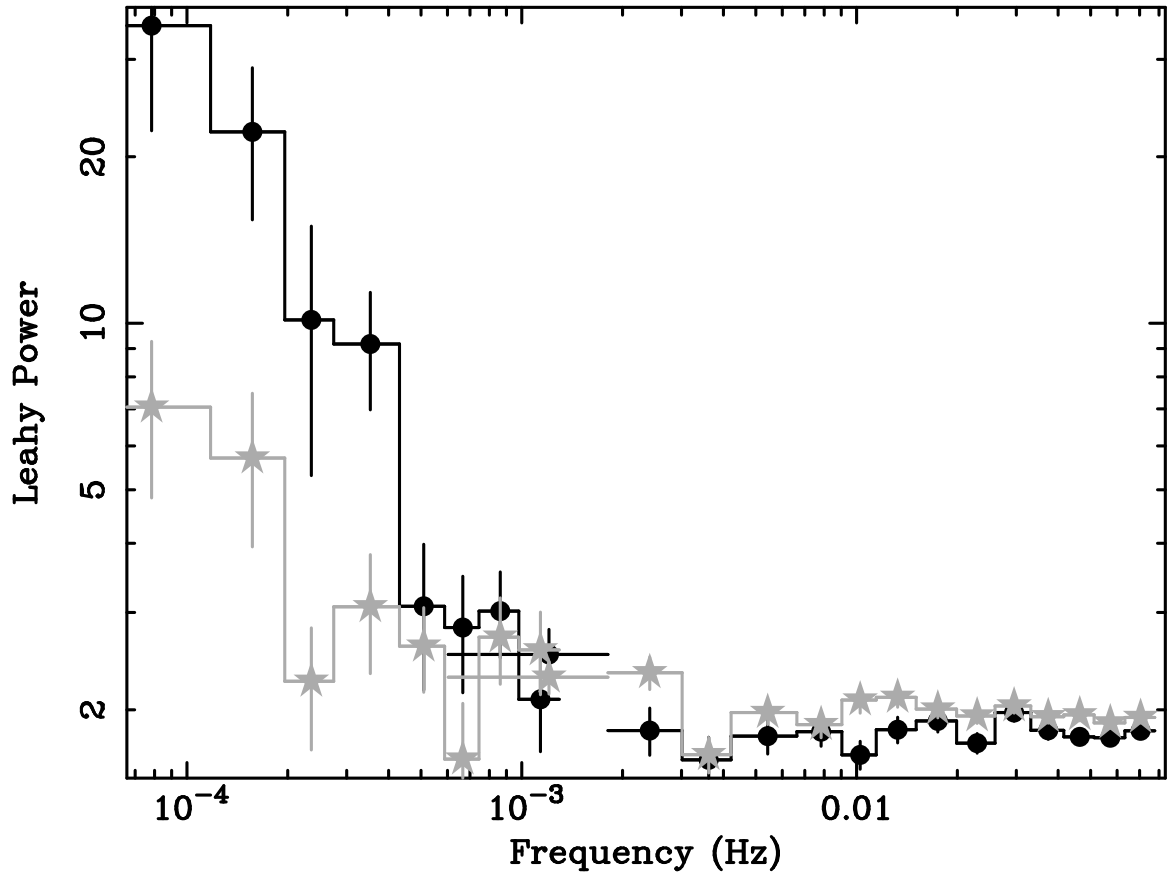}
\caption{Power Density Spectrum of \mulx\ in 2000 March(black) and 2004 July
(grey), normalized according to the prescription of \citet{Lea1983} so that
the expected Poisson noise results in a mean power of 2.}
\label{pdslong}
\end{figure}

\begin{figure}
\plotone{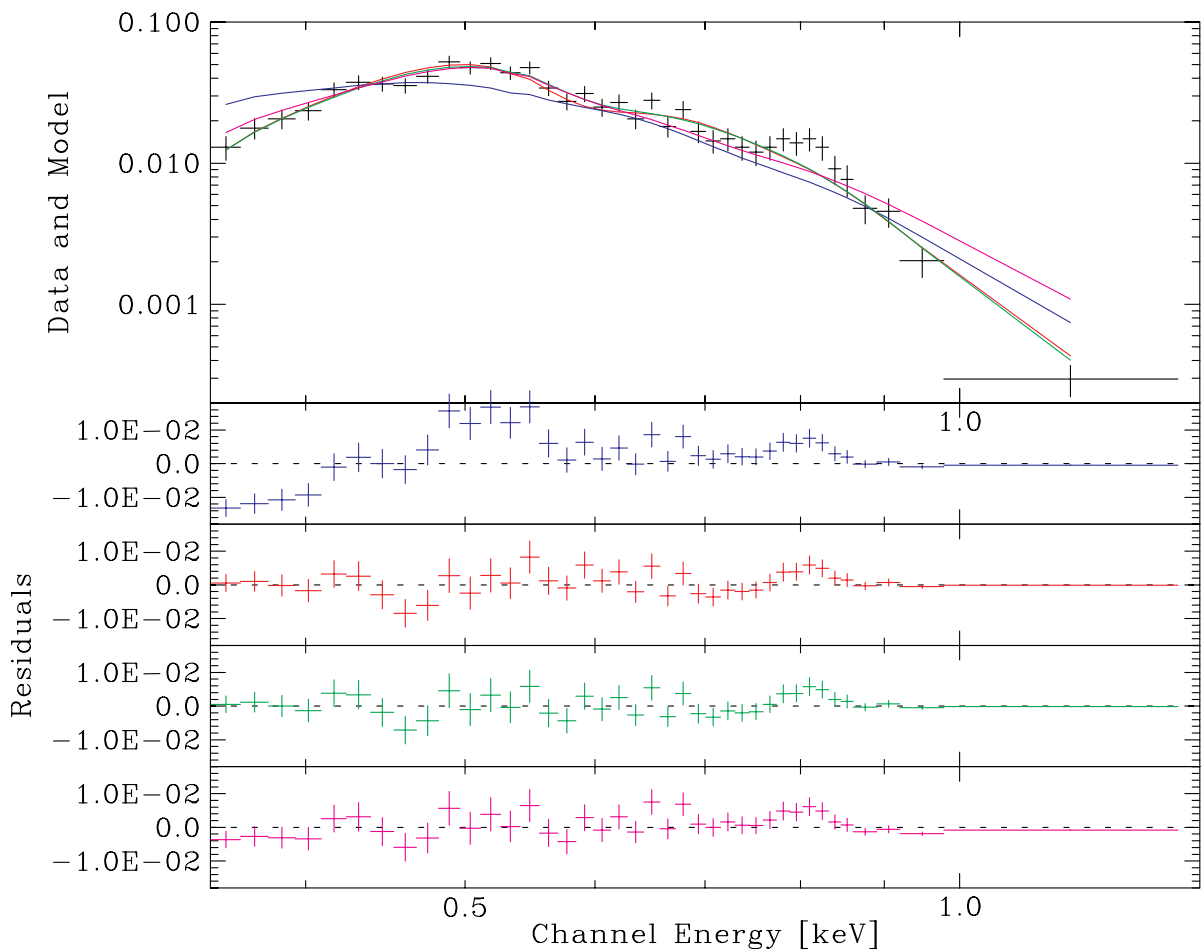}
\caption{The average \chandra\ spectrum of \mulx\ in 2004 July,
as observed, with best-fit blackbody with a $\xi=1$ absorber model (blue),
a blackbody with a neutral absorber model (red), a blackbody model with
edges (green), and a blackbody plus diskline model (magenta).  Residuals
are shown for these models in the lower panels.}
\label{absori}
\end{figure}

\begin{figure}
\plotone{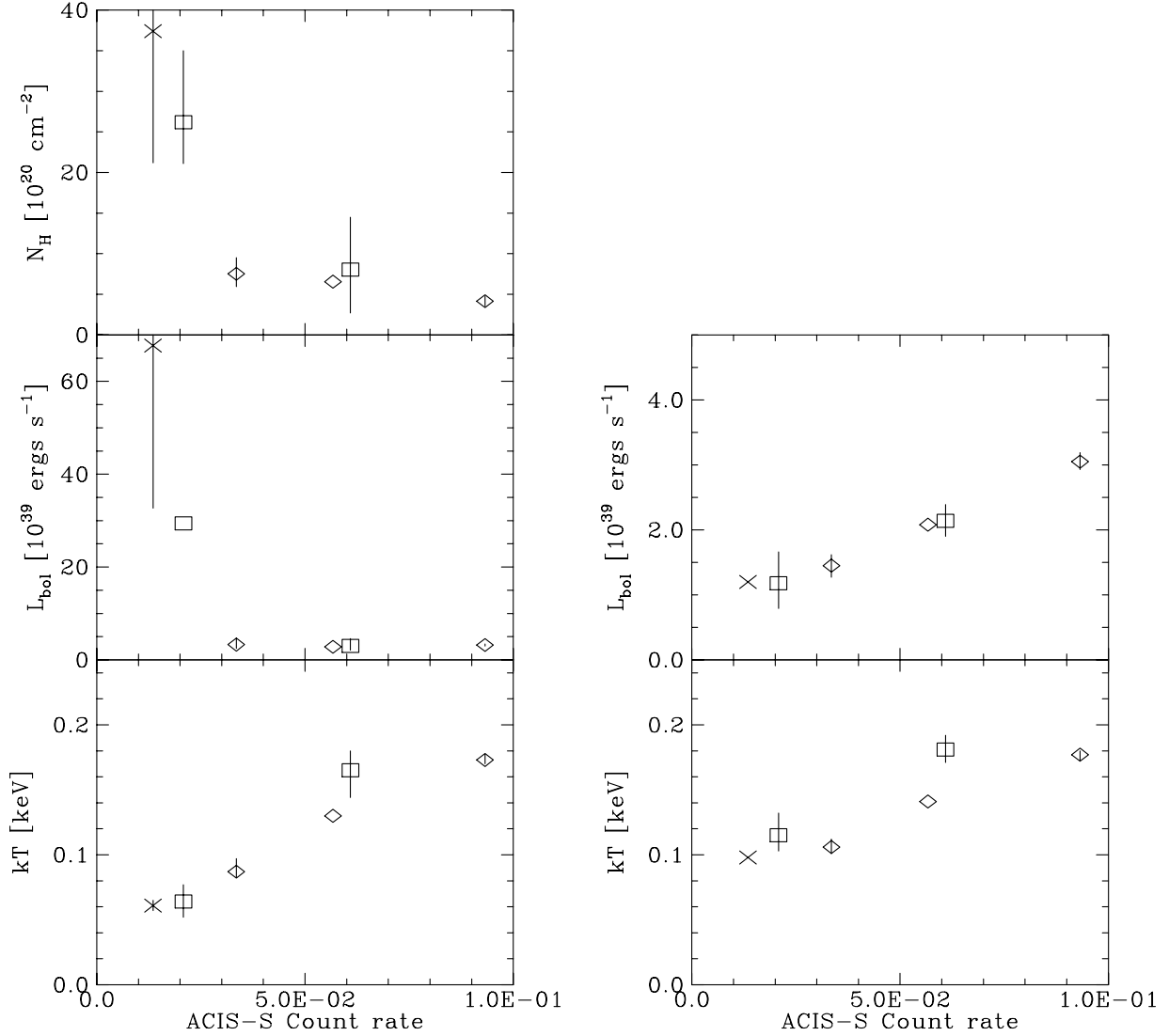}
\caption{The spectral parameters plotted against scaled ACIS-S count rates
using absorbed blackbody (left) and blackbody plus diskline (right)
models.  Symbols are diamonds for Cycle 1 data (2000 March),
crosses for the combined 2004 July data, and open squares for the two
Megasecond 7 pointings in 2004 December and 2005 January.  For those
with $\chi^2_\nu<2$, error bars are shown, but those for \nh\ and L$_{\rm bol}$
for 2004 July data on the left are truncated.
See text for details of the spectral models.}
\label{trend}
\end{figure}

The long-term light curve suggests that \mulx\ is usually in a low
state, with secular average luminosity of about 10$^{37}$ \eps.
For a black hole with accretion efficiency (L/$\dot m c^2$)
of 10\%, an accretion rate of 1.1$\times 10^{17}$ g\,s$^{-1}$ or
1.8$\times 10^{-9}$ M$_\odot$\,yr$^{-1}$ is required to produce
10$^{37}$ \eps.  It is possible to achieve such an accretion rate
through wind accretion, without the mass donor filling its Roche lobe.
If so, the long-term variation may reflect the orbital period of an
eccentric binary.  For example, GX301$-$2 is a 41.5 day period, eccentric
($e$=0.462) binary in which a neutron star accretes the wind of
a supergiant mass donor (see, e.g., \citealt{GX301}).  The analogy
would be very close if the long-term modulation is an irregular
manifestation of an underlying 45-day clock (Figure~\ref{periods} right).

In conclusion, the observations so far of \mulx\ are consistent
with this being a high mass X-ray binary with a B supergiant
mass donor and a large stellar mass black hole.  The long-term
variability may have an underlying periodic clock, and this should
be investigated with further X-ray observations.  At the same time,
ground-based optical spectroscopy should be pursued in search of
the orbital period.





\clearpage

\begin{deluxetable}{lrllll}
\tablecaption{X-ray Observations of \mulx.\label{hsdata}}
\tablehead{\colhead{Observatory} & \colhead{Observation ID} &
           \colhead{Date} & \colhead{MJD} &
           \colhead{Exposure (ksec)\tablenotemark{a}} & \colhead{Notes} }

\startdata

\rosat   & RP600108n00 & 1991-06-08 & 48416 &  33.9 ksec & PSPCB \\
\rosat   & RH600092n00 & 1992-01-09 & 48631 &  18.6 ksec & HRI \\
\rosat   & RH600383n00 & 1992-12-10 & 48967 &  32.6 ksec & HRI \\
\rosat   & RH600820n00 & 1996-05-14 & 50218 & 108.9 ksec & HRI \\
\rosat   & RH600820a01 & 1996-11-21 & 50409 &  68.9 ksec & HRI \\
\chandra &         934 & 2000-03-26 & 51629 &  99.5 ksec & High state\tablenotemark{b} \\
\chandra &        2065 & 2000-10-29 & 51846 &   9.8 ksec & \\
\xmm     &  0104260101 & 2002-06-04 & 52439 & 13.9 ksec/21.5 ksec & \\
\chandra &        4731 & 2004-01-19 & 53023 &  57.0 ksec & \\
\chandra &        5296 & 2004-01-21 & 53025 &   3.2 ksec & \\
\chandra &        5297 & 2004-01-24 & 53028 &  22.0 ksec & \\
\chandra &        5300 & 2004-03-07 & 53071 &  52.8 ksec & \\
\chandra &        5309 & 2004-03-14 & 53078 &  71.7 ksec & \\
\chandra &        4732 & 2004-03-19 & 53083 &  70.7 ksec & \\
\chandra &        5322 & 2004-05-03 & 53128 &  65.5 ksec & \\
\chandra &        4733 & 2004-05-07 & 53133 &  25.1 ksec & \\
\chandra &        5323 & 2004-05-09 & 53134 &  43.2 ksec & \\
\chandra &        5337 & 2004-07-05 & 53191 &  10.1 ksec & High state\tablenotemark{b} \\
\chandra &        5338 & 2004-07-06 & 53192 &  28.9 ksec & High state\tablenotemark{b} \\
\chandra &        5239 & 2004-07-07 & 53193 &  14.5 ksec & High state\tablenotemark{b} \\
\chandra &        5340 & 2004-07-08 & 53194 &  55.1 ksec & High state\tablenotemark{b} \\
\chandra &        4734 & 2004-07-11 & 53197 &  35.9 ksec & High state\tablenotemark{b} \\
\xmm     &  0164560701 & 2004-07-23 & 53210 &  15.1 ksec/21.5 ksec & \\
\chandra &        6114 & 2004-09-05 & 53254 &  67.1 ksec & \\
\chandra &        6115 & 2004-09-08 & 53257 &  36.2 ksec & \\
\chandra &        6118 & 2004-09-11 & 53260 &  11.6 ksec & \\
\chandra &        4735 & 2004-09-12 & 53261 &  29.2 ksec & \\
\chandra &        4736 & 2004-11-01 & 53311 &  78.3 ksec & \\
\chandra &        6152 & 2004-11-07 & 53316 &  44.7 ksec & \\
\chandra &        6170 & 2004-12-22 & 53361 &  48.6 ksec & \\
\chandra &        6175 & 2004-12-24 & 53364 &  41.2 ksec & \\
\chandra &        6169 & 2004-12-30 & 53369 &  29.8 ksec & High state\tablenotemark{b} \\
\chandra &        4737 & 2005-01-01 & 53372 &  22.1 ksec & High state\tablenotemark{b} \\
\xmm     &  0212480201 & 2005-01-08 & 53379 & 9.7 ksec/16.0 ksec & High state\tablenotemark{b} \\

\enddata

\tablenotetext{a}{For \xmm\ observations, the first number is the EPIC-pn
                  exposure time and the second is the EPIC-MOS exposure.}
\tablenotetext{a}{Denotes observations used in our spectral analysis of
                  the high state (there are other observations in which
		  the source was in a high state, but did not warrant
		  spectral analysis).  All 5 sequences taken in 2004 July
		  were combined in our analysis.}

\end{deluxetable}

\begin{deluxetable}{llll}
\tablecaption{Ionized Absorber Fit for 2004 July data.\label{absfit}}
\tablehead{\colhead{\nh} & \colhead{$\xi$}
	& \colhead{kT} & \colhead{$\chi^2_\nu$/d.o.f} \\
           \colhead{($10^{21}$ cm$^{-2}$)} & & \colhead{(eV)} & }

\startdata
$3.3^{+0.7}_{-0.5}$ & 0.0 ($< 10^{-4}$) & 62.4$^{+2.0}_{-5.9}$ & 1.3/35 \\
1.4\tablenotemark{a} & 1.0 (fixed) & 89.6\tablenotemark{a} & 5.7/36 \\
\enddata

\tablenotetext{a}{Errors not calculated because of the poorness of the fit.}

\end{deluxetable}

\begin{deluxetable}{llllllllll}
\tablecaption{Absorbed Blackbody vs. Blackbody plus Disk Line Fits.\label{trendtab}}
\tablehead{\colhead{Data} & \multicolumn{4}{c}{Absorbed Blackbody Fit} &
           \multicolumn{5}{c}{Blackbody plus Disk Line Fit} \\
	    & \colhead{$\chi^2_\nu$} & \colhead{N$_{\rm H}$}\tablenotemark{a} &
	   \colhead{kT} & \colhead{L$_{bol}$}\tablenotemark{b} &
	   \colhead{$\chi^2_\nu$} & \colhead{kT} &
	   \colhead{Line}\tablenotemark{c} &
	   \colhead{L$_{bol}$}\tablenotemark{d} & \colhead{Eq.W} \\
	    & & & \colhead{(eV)} & & & \colhead{(eV)} & & & \colhead{(eV)} }

\startdata
2000 Mar Low      & 1.97 & 7.5 & 89.6 & 3.3 & 1.71 & 106.3 & 3.72 & 1.45 & 102 \\
2000 Mar Medium   & 2.60 & 6.6 & 129.9 & 2.8 & 3.08 & 140.9 & 1.39 & 2.08 & 24\\
2000 Mar High     & 1.59 & 4.2 & 173.1 & 3.2 & 1.55 & 177.1 & 1.32 & 3.05 & 18 \\
2004 Jul          & 1.15 & 32.8 & 61.2 & 67.7 & 1.63 & 98.18 & 3.56 & 1.20 & 125\\
2004 Dec          & 0.65 & 37.4 & 64.3 & 29.4 & 1.08 & 115.0 & 5.16 & 1.18 & 164\\
2005 Jan          & 0.80 & 8.0 & 161.8 & 3.0 & 0.77 & 181.1 & 2.65 & 2.14 & 53 \\
2005 Jan XMM      & 1.85 & 25.6 & 54.74 & 55.8 & 1.36 & 65.26 & 3.63 & 2.25 & 212 \\
\enddata

\tablenotetext{a}{Fitted column density in 10$^{20}$ cm$^{-2}$.}
\tablenotetext{b}{Inferred bolometric luminosity of the blackbody
                  in 10$^{39}$ \eps.}
\tablenotetext{c}{Fitted line flux in 10$^{-5}$ photons cm$^{-2}$s$^{-1}$}
\tablenotetext{b}{Inferred bolometric luminosity of the blackbody
                  plus that of the line in 10$^{39}$ \eps.}

\end{deluxetable}


\clearpage

\begin{thebibliography}{}

\bibitem[Barnard et al. (2003)]{Bea2003} Barnard, R., Osborne, J.P.,
        Kolb, U. \& Borozdin, K.N. 2003, A\&A, 405, 505

\bibitem[Barnard et al. (2004)]{Bea2004} Barnard, R., Kolb, U. \&
        Osborne, J.P. 2004, A\&A, 423, 147

\bibitem[Bohlin et al. (1978)]{Bea1978} Bohlin, R.C., Savage, B.D.
        \& Drake, J.F. 1978, ApJ, 224, 132

\bibitem[Branduardi-Raymont et al. (2001)]{Bea2001} Branduardi-Raymont, G.,
        Sako, M., Kahn, S.M., Brinkman, A.C., Kaastra, J.S. \& Page, M. 2001,
	A\&A, 365, L140

\bibitem[Clarkson et al. (2004)]{Cea2004} Clarkson, W.I., Charles, P.A.,
        Coe, M.J. \& Laycock, S. 2004, MNRAS, 343, 1213

\bibitem[Di Stefano \& Kong (2003)]{dSK2003} Di Stefano, R. \& Kong, A.K.H.
        2003, ApJ 592, 884

\bibitem[Di Stefano \& Kong (2004)]{dSK2004} Di Stefano, R. \& Kong, A.K.H.
        2004, ApJ 609, 710

\bibitem[Di Stefano et al. (2004)]{DiSea2004} Di Stefano, R., Kong, A.K.H.,
        Greiner, J., Primini, F.A., Garcia, M.R., Barmby, P., Massey, P.,
	Hodge, P.W., Williams, B.F., Murray, S.S., Curry, S. \& Russo, T.A.
	2004, ApJ, 610, 247

\bibitem[Fabbiano (2005)]{F2005} Fabbiano, G. 2005, Science, 307, 533.

\bibitem[Gleissner et al. (2004a)]{Gea2004a} Gleissner, T., Wilms, J.,
        Pottschmidt, K., Uttley, P., Nowak, M.A. \& Staubert, R. 2004,
	A\&A, 414, 1091

\bibitem[Gleissner et al. (2004b)]{Gea2004b} Gleissner, T., Wilms, J.,
        Pooley, G.G., Nowak, M.A., Pottschmidt, K., Markoff, S., Heinz, S.
        Klein-Wolt, M., Fender, R.P. \& Staubert, R. 2004, A\&A, 425, 1061


\bibitem[Hickox et al. (2004)]{Hea2004} Hickox, R.C., Narayan, R.
        \& Kallman, T.R. 2004, ApJ 614, 881

\bibitem[Jenkins et al. (2005)]{Jea2005} Jenkins, L.P., Roberts, T.P.,
        Warwick, R.S., Kilgard, R.E. \& Ward, M.J. 2005, MNRAS 357, 401

\bibitem[Kahabka \& van den Heuvel (1997)]{KvdH1997}
        Kahabka, P. \& van den Heuvel, E.P.J. 1997, Ann. Rev. A. Ap. 33, 69

\bibitem[Kato (1997)]{K1997} Kato, M. 1997, ApJS, 113, 121

\bibitem[Kong et al. (2004)]{KDY2004} Kong, A.K.H., Di Stefano, R.
        \& Yuan, F. 2004, ApJ, 617, L49

\bibitem[Kreykenbohm et al. (2004)]{GX301} Kreykenbohm, I., Wilms, J.,
        Coburn, W., Kuster, M., Rothschild, R.E., Heindl, W.A., Kretschmar,
	P. \& Staubert, R. 2004, A\&A, 427, 975

\bibitem[Kuntz (2005)]{K2005} Kuntz, K.D. 2005, ApJ, in preparation

\bibitem[Kuntz et al. (2005)]{Kea2005} Kuntz, K.D., Gruendl, R.A.,
        Chu, Y.-H., Chen, C.-H. R., Still, M., Mukai, K. \& Mushotzky, R.F.
	2005, ApJ, 620, L31

\bibitem[Lasota (2001)]{L2001} Lasota, J.-P. 2001, NewAR, 45, 449

\bibitem[Leahy et al. (1983)]{Lea1983} Leahy, D.A., Darbro, W., Elsner, R.F.,
        Weisskopf, M.C., Kahn, S., Sutherland, P.G. \& Grindlay, J.E.
	1983, ApJ, 266, 160

\bibitem[Miyamoto et al. (1995)]{Mea1995} Miyamoto, S., Kitamoto, S.,
        Hayashida, K. \& Egoshi, W. 1995, ApJ, 442, L13

\bibitem[Mukai et al. (2003)]{Mea2003} Mukai, K., Pence, W.D., Snowden, S.L.
        Kuntz, K.D. 2003, ApJ, 582, 184

\bibitem[Pence et al. (2001)]{Pea2001} Pence, W.D., Snowden, S.L.,
        Mukai, K. \& Kuntz, K.D. 2001, ApJ, 561, 189

\bibitem[Stetson et al. (1998)]{Sea1998} Stetson, P.B., et al. 1998,
        ApJ 508, 491

\bibitem[van der Klis (1994)]{vdK1994} van der Klis, M. 1994, ApJS, 92, 511

\bibitem[Wojdowski et al. (2000)]{Wea2000} Wojdowski, P.S., Clark, G.W.
        \& Kallman, T.R. 2000, ApJ 541, 963

\end{thebibliography}
\end{document}